\documentclass[runningheads]{llncs}
\usepackage[T1]{fontenc}
\usepackage{graphicx}
\begin{document}
\title{Fuzzy synthetic method for evaluating explanations in recommender systems}
%
%
\author{Jinfeng Zhong\inst{1} \and
Elsa Negre\inst{2}}
\authorrunning{Zhong and Negre}
%
\institute{Paris-Dauphine University, PSL Research
Universities, CEREMADE, 75016 Paris, France \email{jinfeng.zhong@dauphine.eu}\and
Paris-Dauphine University, PSL Research
Universities, LAMSADE, 75016 Paris, France
\email{elsa.negre@lamsade.dauphine.fr}}
\maketitle              
\begin{abstract}
Recommender systems aim to help users find relevant items more quickly by providing personalized recommendations. Explanations in recommender systems help users understand why such recommendations have been generated, which in turn makes the system more transparent and promotes users' trust and satisfaction. In recent years, explaining recommendations has drawn increasing attention from both academia and from industry. In this paper, we present a user study to investigate context-aware explanations in recommender systems. In particular, we build a web-based questionnaire that is able to interact with users: generating and explaining recommendations. With this questionnaire, we investigate the effects of context-aware explanations in terms of efficiency, effectiveness, persuasiveness, satisfaction, trust and transparency through a user study. Besides, we propose a novel method based on fuzzy synthetic evaluation for aggregating these metrics. 

\keywords{Evaluating explanation  \and Fuzzy synthetic method \and Explainable recommendation.}
\end{abstract}
\section{Introduction}
Recommender systems (RSs) have become ubiquitous in various online platforms, aiding users in navigating vast amounts of content to find items that meet their preferences or needs. These systems, which power suggestions on platforms ranging from e-commerce to multimedia services, leverage user data to predict and recommend items. As the influence of RSs grows, so too does the necessity for these systems to not only be effective but also transparent and understandable to users. This necessity has spurred the development of explainable RSs, which aim to enhance user trust by providing clear and logical justifications for their recommendations.

The body of literature on RSs has extensively discussed the benefits of explainability. Tintarev and Masthoff \cite{tintarev2015explaining} underscore the importance of explanations in RSs, noting the potential aims of explanations for recommender systems (RSs) include efficiency, effectiveness, persuasiveness, satisfaction, trust and transparency, more details about these terms are presented in Table~\ref{tab:Explanation}. These attributes are essential as they not only help in aligning the system's functionality with user expectations but also enhance the user's control over the decision-making process. However, evaluating the effectiveness of these explanations presents a complex challenge, primarily due to the multifaceted nature of explanation quality itself. Research in this domain has identified several key metrics for assessing the quality of explanations, including transparency, effectiveness, persuasiveness, efficiency, satisfaction, and scrutability (\cite{bilgic2005explaining,gedikli2014should}. Each of these metrics taps into different aspects of the user experience, making it difficult to aggregate them into a unified measure of explanation quality. The challenge lies in the trade-offs that may occur when optimizing one aspect at the expense of another or that these metrics ma correlate with eah other. For instance, an explanation that is very detailed (hence more transparent) might be less persuasive if the user perceives it as too complex or verbose. Therefore, a key research question is how to aggregate the metrics to get an overall evaluation of explanations.

Amidst these challenges, our research contributes to the existing body of knowledge from three aspects. First, we have developed and deployed a web-based questionnaire that facilitates real-time interaction between users and the RS. This tool not only allows us to collect granular data on user preferences and reactions to different types of explanations but also simulates a realistic scenario in which users routinely interact with RSs. Second, our study is among the few to compare the impacts of context-free and context-aware explanations within RSs. By integrating contextual information into the explanations, we investigate whether and how this influences the perceived relevance and effectiveness of the recommendations provided. Third, we propose the use of a fuzzy synthetic method \cite{lan2005decision} for aggregating the various metrics used to evaluate explanations. This method allows for a more comprehensive and nuanced understanding of how different explanation attributes contribute to overall user satisfaction and system effectiveness.

By employing these methodologies within a structured experimental framework using the CoMoDa dataset \cite{kovsir2011database}, our study not only assesses the direct effects of different explanation styles on user experience but also demonstrates the practical utility of the fuzzy synthetic approach in real-world RS applications. Through this threefold contribution, we provide a robust framework for enhancing the transparency, effectiveness, and user-centricity of RSs, thus bridging the gap between technical algorithmic performance and practical user engagement. The remainder of the paper is structured as follows. In Section~\ref{Sec:Rep}, we review start-fo-art related to explanations in recommendations, the metrics of evaluating explanations in RSs. In Section~\ref{sec:ourpro}, we present in detail the problem formulation and the user study design. In Section~\ref{Sec:Exp}, we present the data and analysis obtained from the user study. Lastly, we conclude and propose potential future works.

\begin{table}[t]
\centering

\label{tab:Explanation}\centering
\caption{Potential goals of explanations in recommender system}
\begin{tabular*}{12cm}{ll}
\hline
 Aims& Descriptions  \\
\hline
Efficiency  & Help users make decisions more quickly \\
Effectiveness  & Helps users make better decisions\\
Persuasiveness  & Change users' behaviours  \\
Trust  & Increase user's confidence in the system\\
Transparency  & Help users understand how the system works\\
Satisfaction  &Increase the ease of use or enjoyment \\
\hline
\end{tabular*}

\end{table}
%
%
%

\section{Related work}\label{Sec:Rep}

\subsection{Explanations in recommender systems} \label{sec:expinrec}
Recommender techniques help people find items of their interests faster. As the deployment of these techniques have become ubiquitous, even for high stake decisions, such as medical care and financial services. Users may require explanations for recommendations to help them make better decisions, more precisely, help them evaluate the quality of the recommended items \cite{tintarev2022beyond}. Appropriate explanations can boost users' trust and satisfaction, which is proved by numerous former studies \cite{herlocker2000explaining,bilgic2005explaining}.  

Depending on the information included in the explanations, state-of-art explanations in RSs can be classified into similar user (item) explanation, feature-based explanation, opinion-based explanation and social explanation \cite{zhang2020explainable}. However, these explanations do not consider users' contextual situations. Therefore, they may not satisfy users' needs under different situations. To the best of our knowledge, few studies have combined CARSs and explanations in RSs. State-of-art work on CARSs focuses on better modeling users' preferences under different situations, however users' contextual situations that can influence their preferences have not been sufficiently explored to generate explanations. The effects of context-aware explanations have not been fully studied yet. Baltrunas et al. \cite{baltrunas2011context} showed that context-aware explanations can promote user satisfaction when recommending places of interest.  Recently, Sato et al. \cite{sato2018explaining} proposed context style explanation in restaurant recommendation. They showed through a user study that context-aware explanation can promote persuasiveness and usefulness.  However, none of these works has fully explored the effects of context-aware explanations in RSs in terms of efficiency, effectiveness, persuasiveness, transparency, satisfaction, trust. We believe that CARSs and explanations in RSs are two rich areas that can actually be combined to generate context-aware explanations, which is one of the motivation of our work.

\subsection{Metrics}\label{Sec:Metr}

As asserted by \cite{tintarev2022beyond}, finding the metrics right is a key part of any evaluation. It should be noted that although the term ``explanation" has been studied by researchers from different domains, there is no standard definition of ``what is explanation". According to \cite{tintarev2022beyond}, explanations in RSs are descriptions that help users understand the quality of items. This definition is to some extent abstract and difficult to evaluate. Therefore, we decided to evaluate explanations through possible aims of explanations defined by \cite{tintarev2015explaining} as presented in Table~\ref{tab:Explanation}. In this work, the six aims serve as metrics for evaluating explanations. In this subsection, we will introduce methods for evaluating the six aims presented in Table~\ref{tab:Explanation}.

As defined by \cite{tintarev2015explaining}, an efficient explanation helps users to decide the best recommended item faster. Efficiency is evaluated differently in different application domains. In knowledge-based RSs proposed by \cite{pu2006trust}, explanations can be generated by showing the trade-off between item properties, efficiency of explanations can therefore be measured by the number of interactions needed for users to make a decision. In conversational RSs, efficiency can be measured by the total number of dialogue steps \cite{thompson2004personalized} or time \cite{mccarthy2005experiments} required before a recommendation is accepted. In other cases, efficiency is usually measured by the time consumed for a decision to be made by users \cite{gedikli2014should}, for example the time needed to select an item or to give a rating. Since explaining is a human-centered process, it would also be helpful measure whether explanations are perceived to be efficient. To the best of our knowledge, former works did not report the results of human perceived efficiency.

Effectiveness measures how well an explanation helps users make better decisions \cite{tintarev2015explaining}. In the context of RSs, better decisions mean helping users better understand the quality of recommended items to find relevant ones. According to \cite{tintarev2022beyond}, metrics for evaluating effectiveness of explanation can be classified by: perceived effectiveness and objective measuring. Perceived effectiveness is usually measured by self-report of users \cite{vig2009tagsplanations}. Objective measuring measures the differences between user estimated quality of items and their actual quality, the smaller the differences are, the more effective the explanations are. This is the method adopted in \cite{bilgic2005explaining,mccarthy2005experiments}.

Persuasiveness of explanations measures how well explanation enables RSs to convince participants to make certain decisions \cite{tintarev2015explaining}. Like evaluating effectiveness, evaluation of persuasiveness can be done in two ways: perceived persuasiveness and objective measuring. Herlocker et al.\cite{herlocker2000explaining} studied the persuasiveness of different explanation interfaces by showing 21 interfaces and asking participants directly how likely they were to watch a movie. It can also be approximated by the difference of two ratings: the first being a rating by users given only an explanation and the second being the rating when presented detailed information of the item, this is the method adopted by \cite{bilgic2005explaining,gedikli2014should}.

State-of-art methods for evaluating satisfaction of users mainly include: (1) asking them directly whether a recommender system is nice to use \cite{herlocker2000explaining} or simply whether they are satisfied with the explanations provided; (2) measuring user loyalty (the likelihood of reusing a system) \cite{cramer2008effects}. The first method is more often adopted as the second requires a long term experiment.

Transparency in RSs contains two aspects \cite{gedikli2014should}: (1) objective transparency, which means the actual algorithm of the system; (2) user-perceived transparency, to what degree users think a system is transparent. Presenting user-oriented explanations rather than actual algorithms might be more appropriate \cite{vig2009tagsplanations}. Reasons are summarized as below: algorithms may be too complicated for users to understand; algorithms are intellectual properties that should be protected. For these reasons, transparency is usually evaluated by questionnaires where participants are asked directly to what degree they think a system is transparent when accompanied with the explanations to be evaluated. 

Trust is defined as users' perceived confidence in RSs \cite{chen2005trust}, therefore, a questionnaire is usually adopted to evaluate to what degree participants trust a system. Like the evaluation of satisfaction, trust can be reflected by user loyalty \cite{mcnee2003interfaces}, which requires a long term experiment. 
\section{Our proposition}\label{sec:ourpro}

\subsection{Recommendation method}

We design a web-based questionnaire that is able to interact with participants: recommending and explaining recommendations. To generate context-aware recommendations, we applied a method we proposed in \cite{zhong2022towards1}. The main steps mainly include: (1) Cluster items (movies) using Hierarchical Clustering as \cite{ferdousi2018cbpf}; (2) Calculate the item-based Pearson Correlation Coefficient (PCC) \cite{benesty2009pearson} between each contextual  condition and ratings for each item cluster; (3) Represent each contextual condition based on the calculated PCC; (4) Calculate the similarity between the target contextual situation $CS^*$ and the contextual situations $CS$ existing in the  original dataset; (5) Select a local dataset whose contextual situations are similar to the target contextual situation according to a similarity threshold; (6) A traditional 2D recommender technique is applied in the local dataset to obtain a set of recommendations. For more details, we refer to \cite{zhong2022towards1}.

\section{Experiment setup} \label{Sec:Pro}

We adopt the widely used contextual movie recommendation dataset CoMoDa \cite{kovsir2011database}, where there are 121 users, 1197 movies, 2296 ratings and 12 contextual factors. In the CoMoDa dataset, the average number of ratings of each user is about 12. In order to be consistent with the original dataset, participants are required to give their ratings for 12 movies under different contextual situations. In the original dataset, there are 12 contextual factors, some contextual factors could have a more important impact than others \cite{Zahra2020}. We wish to select the most impacting contextual factors to minimize the efforts of participants to specify his/her contextual situation. According to \cite{Zahra2020}, in the CoMoDa dataset, the 6 most impacting contextual factors are \emph{(i) whether users have watched a movie before, (ii) users are given a decision or decide themselves, (iii) users' physical wellness, (iv) users' moods, (v) users' location and (vi) weather}. Note that in step 2 of our protocol, we recommend movies that participants have not interacted with in step 1 of our protocol, we do not consider the contextual factor \emph{Interaction}. Since all the participants are given a movie, we do not consider the contextual factor \emph{Decision} either. Therefore, in our case, we only consider the following 4 contextual factors: \emph{Physical wellness}, \emph{Mood}, \emph{Location} and \emph{Weather}.

\begin{figure}[t]
	\centering
	\includegraphics[scale=0.35]{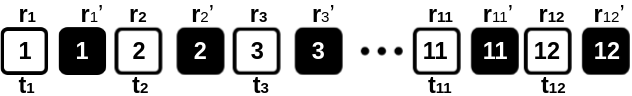}
	\caption{The number indicates the movie recommended. White indicates that only explanation is displayed, black indicates that detailed information of the movie is displayed, $r$ is the rating given by participants, and $t$ is the time cost by participants to give a rating.}
	\label{fig:step}
\end{figure}

There are mainly four steps in the user study\footnote[2]{Here is a demo that shows the procedure of user study: \url{https://drive.google.com/drive/folders/1cYiJ8gLPYlMZu4k41gIBqnE2HytbK3KL?usp=sharing}}: (1) participants are asked to rate 12 randomly selected movies given several contextual situations by applying the data set \cite{kovsir2011database}. This step is to learn users' preferences under different contextual situations; (2) the system generates recommendations for the participants under a randomly generated contextual situation, for more details we refer to \cite{zhong2022towards1}; (3) participants are presented with explanations but without detailed information of movies and they are asked to rate the movie, noted as $r$, the time consumed for giving the rating is noted as $t$, users are then presented with detailed information of the same movie and rate them again, noted as $r^{'}$,  Figure~\ref{fig:step} illustrates this process; (4) participants are asked to evaluate the explanations in terms of transparency, effectiveness, efficiency, persuasiveness, trust, and satisfaction by a 1 (lowest) to 5 (highest) rating scale. In all, this user study follows a within-subject design \cite{keren2014between} to get more data from participants. Therefore, the six metrics can be quantified as below: (1) efficiency: the time used by participants for giving a rating when only presented with explanations; (2) effectiveness: the difference between $r$ and $r^{'}$, if the difference is near 0, it indicates good effectiveness since explanation has helped participants perfectly estimate the rating of recommendation; (3) persuasiveness: $r-r^{'} > 0$ means positive persuasiveness; $r-r^{'} < 0$
means negative persuasiveness; (4) transparency, satisfaction, and trust: they are reflected by users’ responses in step 4;
(5) in terms of efficiency, effectiveness, and persuasiveness, we will compare the measured effects and human-perceived
effects. 

The protocol adopted in this user study basically originated from \cite{bilgic2005explaining}. The protocol was adopted by \cite{gedikli2014should} where the authors studied how explanation style can influence effects of different explanation styles in terms of efficiency, effectiveness, persuasiveness, transparency, satisfaction. Our protocol varies from \cite{gedikli2014should} in two significant ways:

\begin{itemize}
    \item We assess the effects of explanations in terms of trust, which was not considered by \cite{gedikli2014should}.
    \item For efficiency, effectiveness, and persuasiveness, we ask users to evaluate different styles of explanations concerning these three metrics. This is to identify any discrepancies between the human-perceived experiences and actual effects. In contrast, \cite{gedikli2014should} solely utilized objective methods to measure these three metrics.
\end{itemize}

\subsection{Baselines}   
Concretely, we wish to compare the effects of explanations with and without contextual information. Table~\ref{tab:example} presents the explanation types we plan to show to participants in the experiment. The \emph{Avg} style explanation shows users the average rating of the recommended item; \emph{Per} style explanation presents the distribution of ratings of the recommended item; \emph{Simu} style explanation presents the average ratings of similar users. \emph{Simu}, \emph{Per}, and \emph{Avg} style explanations are among the 21 explanation styles that \cite{herlocker2000explaining} compared in terms of persuasiveness. Since explanations that perform well in one goal may perform poorly in another (trade-off among goals) \cite{tintarev2015explaining} and we investigate six different goals, we do not simply select the top-ranking explanation types proposed by \cite{herlocker2000explaining}. We selected the top, middle, and end of the ranking list of the 21 explanation styles, namely \emph{Simu}, \emph{Per}, and \emph{Avg} respectively. \emph{Simi} style explanation simply says the recommended item is similar to the items a participant has interacted with before,  which is popular in websites such as Amazon \cite{sarwar2001item};  \emph{Content} style explanation presents the characteristics of recommended items that may be preferred by users, it is widely adopted for the movie, book recommendation \cite{vig2009tagsplanations}. \emph{Context-aware} explanation leverages users’ contextual information to explain recommendations \cite{sato2018explaining}. In order that these explanation styles can be compared fairly, they are all in textual forms.

\begin{table}[h]
\centering
\caption{Explanations to be presented.}
\label{tab:example}
\begin{tabular}{l|l}
\hline
 Name&Example  \\
\hline
\emph{Avg}  &The average rating of this movie is 3.8 \\
\emph{Per}  &80 percent of users rate this movie more than 4 \\
\emph{Simu}  & The average rating of users whose preferences are similar to yours is 4.1 \\
\emph{Simi}  &This movie is similar to movies you watched before\\
\emph{Content}  & This is a movie directed (acted) by   \\
\emph{Context-} & \textbf{The system suppose that you would like to watch this movie}\\
    \emph{aware}&\textbf{when it shines, healthy you are at home and in good moods}\\
\hline
\end{tabular}
\end{table}

\section{Results and discussion} \label{Sec:Exp}

Given the web-based nature of our questionnaire, we were able to distribute it to researchers both within our laboratory and externally, as well as to students at our university via mailing lists. The questionnaire was available online for a week, allowing participants to respond anonymously. In total, we received 147 responses.

With respect to demographic breakdown, the genders were evenly distributed with almost half of the respondents identifying as male and the other half as female. The majority of participants fell within the age range of 18 to 55 years; over $66\%$ of them had achieved an education level of Bac+5 (equivalent to a master's degree level) or higher. Close to half of the participants were students (ranging from undergraduate to doctoral candidates), while the remainder were either employees of companies or involved in higher education. This distribution is largely due to the fact that our mailing lists primarily consisted of researchers and students. Over half of the participants reported watching movies regularly, indicating a high level of interest in this medium.

In order to analyze the observed differences between the different explanation types, we first applied analysis of variance \cite{tabachnick2007experimental}. Upon identifying significant differences, we proceeded to employ the Tukey HSD test \cite{abdi2010tukey} to pinpoint the specific discrepancies. All tests were conducted at a $95\%$ confidence level.

\subsection{Objective evaluation}

In this section, we will present the results of objective evaluation: efficiency, effectiveness, and persuasiveness. Table~\ref{tab: objective_evaluation} shows the results of these three metrics.

\begin{table}[t]
\centering
\caption[Efficiency, effectiveness, and persuasiveness of different types of explanations.]{Efficiency, effectiveness, and persuasiveness of different types of explanations (the average values). Efficiency is reflected by the time consumed by participants to give a rating (in seconds); effectiveness is reflected by the difference between $r$ and $r^{'}$ (as presented in Figure 1), the closer it is to 0 the more effective an explanation is; persuasiveness is also measured by the difference between $r$ and $r^{'}$,  $r-r^{'} > 0$ means positive persuasiveness; $r-r^{'} < 0$
means negative persuasiveness. Note that $r \in [1,5]$, $r^{\prime} \in [1,5]$, therefore, $r-r^{'} \in [-4,4]$.}
\label{tab: objective_evaluation}
\begin{tabular}{lcc}
\hline
& Efficiency (seconds) & Effectiveness and persuasiveness \\ \hline
\emph{Avg}                  & 4.78           & 0.30                             \\
\emph{Content}              & 5.37           & \textbf{-0.02}                            \\
\emph{Context-aware}              & \textbf{8.25}           & \textbf{-0.01}                            \\
\emph{Per}                  & 5.17           & 0.11                             \\
\emph{Simi}                 & 5.99           & -0.18                            \\
\emph{Simu}                 & 6.89           & 0.10                             \\ \hline
\end{tabular}
\end{table}

Observations indicate that participants presented with Context-aware explanations tend to take more time to make a decision, with decision-making in our context meaning the assignment of a score (See row Context-aware and column Efficiency (seconds)). While there were no significant differences among the explanation types in terms of $r-r^{'}$, both Content style and Context-aware explanations slightly outperformed others in helping participants estimate the quality of recommendations, as the difference between $r$ and $r^{'}$ was closest to zero for these styles. Conversely, Avg and Simi explanation types led users to underestimate the quality of recommendations. The Per and Simu styles were found to be marginally more persuasive than the other styles, suggesting that the presentation of ratings from similar users can enhance persuasiveness.

\subsection{Subjective evaluation}

We now turn to the results of the subjective evaluation, as assessed via a Likert-scale questionnaire. It is important to note that the factors of efficiency, effectiveness, and persuasiveness were also evaluated by directly soliciting feedback from participants. Table~\ref{tab:subjective_eff} displays the results for participant-reported efficiency, effectiveness, persuasiveness, satisfaction, trust, and transparency. In the subjective evaluation, Simi style explanations were perceived as the most efficient, effective, and persuasive. However, this was not mirrored in the objective evaluation (see Table~\ref{tab: objective_evaluation}). In the objective assessment, Avg style explanations were found to be more efficient and persuasive, while Context-aware and Content style explanations proved to be more effective. These results highlight a discrepancy between the actual effects of the explanations and the participants' perceptions. The reasons behind this divergence and strategies to minimize it warrant further investigation.

\begin{table}[]
\centering
\caption[Subjective evaluation of efficiency, effectiveness, persuasiveness, satisfaction, trust, and transparency]{Subjective evaluation of efficiency, effectiveness, persuasiveness, satisfaction, trust, and transparency. Note that these are the results of the Likert-scale questionnaire, 1-5; the values in the table are the average of the participants.} 
\label{tab:subjective_eff}
\begin{tabular}{lcccccc}

\hline
\multicolumn{1}{c}{} & Efficiency& Effectiveness & Persuasiveness& Satisfaction& Trust & Transparency \\ \hline
\emph{Avg}                  & 3.16           & 2.98          & 3.01 & 3.00           & 2.85          & 3.26          \\
\emph{Content}              & 3.25           & 3.12          & 3.20  & 3.19          & 3.28          & \textbf{3.46}          \\
\emph{Context-aware}              & 2.96           & 3.07          & 3.24 & 3.21           & 3.14          & 3.26          \\
\emph{Per}                  & 3.32           & 3.26          & 3.18   & 3.11           & 3.04          & 3.12        \\
\emph{Simi}                 & \textbf{3.72}           & \textbf{3.47}          & \textbf{3.52} & \textbf{3.50}           & \textbf{3.53}          & 3.42     \\
\emph{Simu}                 & 3.49           & 3.33          & 3.30   & 3.36           & 3.30          & 3.14        \\ \hline
\end{tabular}

\end{table}

Satisfaction, trust, and transparency are evaluated by directly asking participants. The last three columns in Table~\ref{tab:subjective_eff} contain the results of participants-reported satisfaction, trust, and transparency. \emph{Simi} style explanations achieve the highest level of satisfaction and trust, which shows that participants prefer explanations that compare the present recommendation and the items that they consumed before. \emph{Content} style explanations provide information about recommendations and help participants understand how the system works. As a result, \emph{Content} style explanations are perceived to be transparent.

Contrary to our hypothesis, context-aware explanations were not perceived as more satisfying or transparent. A potential reason for this could be related to the nature of our user study. Since the contextual situations were randomly generated, they may not have accurately reflected the real situations of the participants when they were responding to the questionnaire. We asked participants to envision themselves in these artificially-created scenarios, which is a non-trivial task. This could also account for why participants took longer to issue a rating when presented with context-aware explanations, as seen in Table~\ref{tab: objective_evaluation}. 

\subsection{Correlation between different metrics}

Table~\ref{tab:correlation} presents the \emph{Spearman Rank Order Correlation} of metrics for evaluating \emph{Context-aware} explanations, it can be observed that these metrics are not independent. For example, in the \emph{Context-aware} explanation, trust and satisfaction, efficiency and effectiveness are positively correlated. However, transparency and persuasiveness, trust and efficiency are negatively correlated, this leads to the conclusion that it is hard to design explanations that can perform well in all aspects. In order to comprehensively evaluate the qualities of explanations, more metrics should be considered.

\begin{table}[]
\centering
\caption[Spearman Rank Order Correlation of metrics]{Spearman Rank Order Correlation of metrics in \emph{Context-aware} explanations. The correlation is between $-1$ and $+1$, large positive values indicate that two metrics are positively correlated, and vice versa.}
\label{tab:correlation}
\begin{tabular}{ccccccc}
\hline
               & Efficiency & Effectiveness & Persuasiveness & Satisfaction & Trust   & Transparency \\ \hline
Efficiency     & 1.0***     & 0.45***       & 0.05           & -0.17        & -0.15   & 0.01         \\
Effectiveness  & 0.45***    & 1.0***        & 0.01           & -0.08        & -0.0    & -0.1         \\
Persuasiveness & 0.05       & 0.01          & 1.0***         & 0.04         & -0.03   & -0.17        \\
Satisfaction   & -0.17      & -0.08         & 0.04           & 1.0***       & 0.82*** & -0.04        \\
Trust          & -0.15      & -0.0          & -0.03          & 0.82***      & 1.0***  & -0.11        \\
Transparency   & 0.01       & -0.1          & -0.17          & -0.04        & -0.11   & 1.0***       \\ \hline
\end{tabular}

\end{table}

\subsubsection{Aggregating metrics}

\begin{figure}[t]
	\centering
	\includegraphics[scale=0.20]{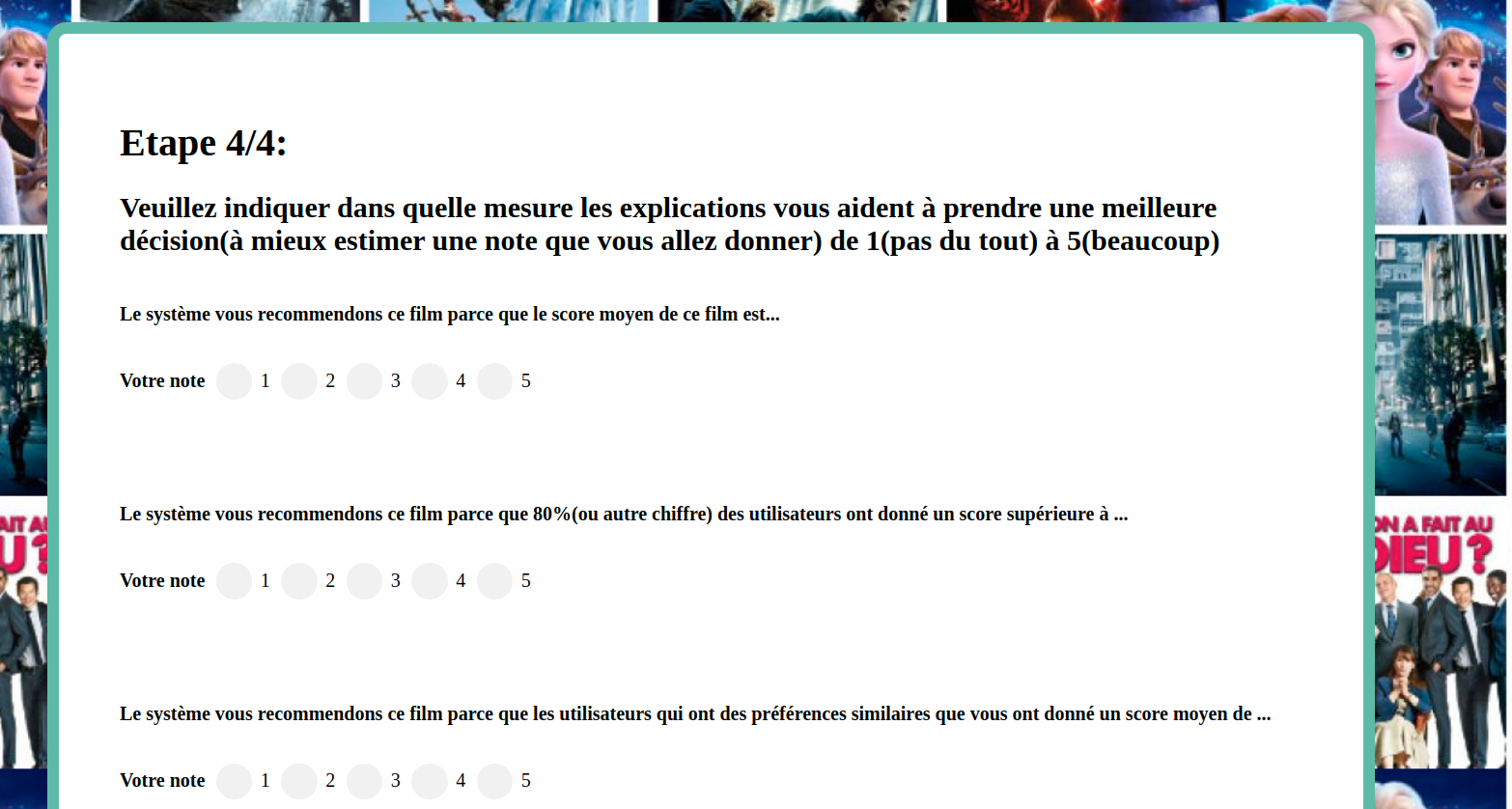}
	\caption[A screenshot of the Likert-scale questionnaire]{A screenshot of the Likert-scale questionnaire when evaluating the effectiveness of explanations. Participants are indicated that 1 means they think the explanation does not help them make good decision at all while 5 indicates that the explanation help them make good decisions.}
	\label{fig:capture}
\end{figure}

The values in Table~\ref{tab:subjective_eff} are human perceived, human perception is always fuzzy and the evaluator’s opinion by nature comes in linguistic form \cite{zadeh2013computing}, which is the reason why we have specified the meaning of each score when participants respond to the Likert-scale questionnaire, see Figure~\ref{fig:capture}. Therefore, the score in our questionnaire can be linked to categorical appraisal grades: ``Very poor'', ``Poor'', ``Medium'', ``Good'' and ``Very good''. On the other hand,  the metrics are somehow correlated as indicated in Table ~\ref{tab:correlation}, aggregating these metrics by applying the average operation is not realistic. To get a more general evaluation of explanations, we are inspired by the  fuzzy synthetic evaluation \cite{lan2005decision} that provides a fuzzy mapping between each of the evaluation factors (e.g. efficiency, effectiveness, persuasiveness, trust, transparency, and  satisfaction in our case) to a set of categorical appraisal grades (``Very poor'', ``Poor'', ``Medium'', ``Good'', ``Very good'' in our case.).  For instance, a score such as 3 on a 5-point scale could be classified under both ``Medium'', ``Good'' grades. The degree to which the score aligns with each grade can vary, embodying different membership degrees to each grade. This variation depends on the weights attributed to each evaluation factor and the average score given by participants. Taking an example, one might discern that a score of $3.2$ for efficiency could correspond to the fuzzy sets ``Medium'', ``Good'' and ``Very Good'', with respective membership degrees of 0.7, 0.2, and 0.1. Assigning a degree of membership to multiple ``fuzzy grades'' allows for capturing and preserving more of the inherent uncertainties involved when responding to the questionnaire. We follow the steps given by \cite{zhou2017using}\footnote[2]{In the paper, the authors propose a fuzzy comprehensive evaluation method to determine product usability.}:

\begin{itemize}
    \item \textbf{Step 1, determining the set of evaluation factors:} evaluation factors can be determined based on the goals of the product evaluation process: $\mathcal{F} = \{f_1,f_2,\dots, f_s\}$ indicates a set of $s$ factors. In our case, $\mathcal{F} = \{efficiency, effectiveness, persuasiveness, trust, transparency, \\ satisfaction\}$.
    \item \textbf{Step 2, determining the set of appraisal grades:} this set defines the levels of appraisal grades, e.g. $\mathcal{V} = \{v_1,v_2,\dots,v_p\}$ defines $p$ levels of appraisal grades. In accordance with the linguistic description in the questionnaire, the appraisal grades set is defined as $\mathcal{V} = \{Very poor, Poor, Medium,Good, Very good\}$. 
    \item \textbf{Step 3, setting the fuzzy mapping matrix:} the aim of the evaluation process is to establish a map from set $\mathcal{F}$ to set $\mathcal{V}$. For a specific factor $f_i$ the fuzzy mapping to the appraisal vector $\mathcal{V}$ can be represented by the vector $r_i = (r_{i1},r_{i2},\dots,r_{ik},\dots,r_{ip})$ where $p$ denotes the number of levels of the appraisal grades in \textbf{Step 2}, $r_{ik}$ denotes the fuzzy membership degree of evaluation factor $i$ to grade $k$. Using the example of efficiency above, $r_1 = (0,0,0.7,0.2,0.1)$,  then the measurement on the evaluation factor ``efficiency'' has a fuzzy membership of $0.7$ in the grade ``Medium'', a fuzzy membership of $0.2$ in the grade ``Good'' and a fuzzy membership of $0.1$ in the grade ``Verygood'', respectively. Therefore, the size of the fuzzy mapping matrix $\mathcal{R}$ is $s \times p$, in our case, the size becomes $6 \times 5$. Intuitively, we set the $r_{ik}$ as the proportion of participants that have given the score corresponding to the appraisal grade level $k$ in terms of evaluation factor $i$. Still in the example of efficiency, if 7 out of 10 participants have given a score of 3 then $r_{13} = 0.7$.
\item \textbf{Step 4, determining the weight of each evaluation factor:}  for a thorough assessment, it's imperative to quantify the relative weight of each evaluation factor on the overall quality. This is particularly relevant as different stakeholders place varying levels of importance on these factors. For instance, service providers, such as online sellers, may place a higher emphasis on the persuasiveness of explanations, whereas users might prioritize the effectiveness of explanations. To this end, the weights can be represented by a vector: $\mathcal{W} = (w_1,w_2,\dots, w_s)$, where $\sum\limits_{i=1}^{i=s}w_i = 1$. To the best of our knowledge, this is the first work that applies fuzzy mapping matrix to aggregate different metrics for evaluating the qualities of explanations. We empirically set the weight of the six metrics the same: $\mathcal{W} = (\frac{1}{6},\frac{1}{6},\frac{1}{6}, \frac{1}{6},\frac{1}{6},\frac{1}{6})$\footnote[3]{Note that the weight could be set differently according to the main goals of explanations. For example, explanations designed to persuade users to buy or to consume, the weight of persuasiveness can be set higher.}.
\item \textbf{Step 5, getting the overall appraisal result:} the final assessment outcome can be computed by considering the relative weights of each evaluation factor. As a result, a singular vector corresponding to the overall evaluation can be expressed as $\mathcal{E} = (e_1,e_2,\dots,e_j,\dots,e_p) = \mathcal{W} \circ \mathcal{R}$, where $\circ$ is the composition operator. There are typically five types of composition operators {zimmermann2011fuzzy, lan2005decision}, we have chosen the sum operator: $e_j = \sum\limits_{i=1}^{i=p} w_i*r_{ij}$. We leave the comparison of the composition operations for future work. 
    
\end{itemize}

Following the outlined steps, the comprehensive evaluation of the six types of explanations is summarized in Table~\ref{tab:fuzzy}. Among the context-free types, the \emph{Simi}, \emph{Content}, \emph{Per}, and \emph{Simu} classifications fall within the ``Good'' level, with the \emph{Simi} type having the highest fuzzy membership. The \emph{Avg} type is situated within the ``Medium'' category. These findings align with Table~\ref{tab:subjective_eff}, where the \emph{Simi} type scores higher than the other types, barring transparency. Interestingly, despite its fuzzy membership standing at just 0.2857, the \emph{Context-aware} type falls within the ``Very Good'' level. We infer that participant evaluations of the \emph{Context-aware} type could be polarized, given the considerable fuzzy membership for ``Very Poor''. This discrepancy could arise if the randomly generated contextual situation aligns closely with the participant's real-life situation, leading them to rate this explanation type more favorably.

\begin{table}[t]
\centering
\caption[Overall evaluation of explanation types]{Overall evaluation of explanation types. Note that, for each type, the largest fuzzy membership is in bold.}
\label{tab:fuzzy}
\begin{tabular}{lc}
\hline
& Overall evaluation  \\ \hline
\emph{Avg}                  & (0.1054,0.1530,\textbf{0.3979},0.2176,0.1258)                                        \\
\emph{Content}              & (0.0544,0.1632,0.3197,\textbf{0.3367},0.1258)                                     \\
\emph{Context-aware}              & (0.1938,0.1667,0.1530,0.2006,\textbf{0.2857})                                 \\
\emph{Per}                  & (0.0748,0.1598.0.3299,\textbf{0.3197},0.1156)                                   \\
\emph{Simi}                 & (0.0442,0.1292,0.2142,\textbf{0.4217},0.1904)                                    \\
\emph{Simu}                 & (0.0714,0.1292,0.3027,\textbf{0.3537},0.1428)                                      \\ \hline
\end{tabular}
\end{table}

\subsection{Discussions}

This primary study presents several limits: first, our context-aware explanations were rather simplistic, merely presenting contextual information. They could, in fact, be combined with item features. Second, the contextual situations were randomly generated, therefore, the contextual information might not always align with the participants' real-life situations. 

Despite these limitations, our user study allows us to derive some valuable insights. First, although context-aware explanations can assist users in better estimating the quality of recommendations, they require more time for users to make a decision. Second, using similar users' ratings to explain recommendations can be persuasive. Third, there may be a discrepancy between the actual effects of explanations and how they are perceived by participants. Lastly, there is an interrelation among the goals of explanations, indicating that it might not be feasible to design explanations that optimize all metrics simultaneously.

\section{Conclusion and perspectives}
In this work, we designed a web-based questionnaire to study the effects of contextual information of explanations in recommender systems (RSs). To the best of our knowledge, this is the first attempt that has fully explored the effects of context-aware explanations in RSs in terms of efficiency, effectiveness, persuasiveness, transparency, satisfaction, trust. We compared Context explanation with other five popular explanations in terms of the six metrics. Results of our experiments indicate that context factors can influence participants' ratings. However, explanations containing only contextual information can not boost the effects of explanations in terms of the six metrics. Our contributions are: (1) We developed a web-based questionnaire that can interact with participants: generating context-aware recommendations and explaining recommendations;  (2) This questionnaire makes it possible to evaluate explanation in terms of efficiency, effectiveness, persuasiveness, satisfaction, transparency and trust;
(3) We apply a fuzzy synthetic method for aggregating different metrics, allowing us to get an overall evaluation of explanations. 

We admit the limitations of our works: (1) In this work, we limited our work in movie recommendation; (2) The situation we used to generate recommendation is not the participants' real situation; (3) We explored explanations containing only contextual information; (4) The participants of the questionnaire are mostly students or employee of higher education. However, we believe that this work is an important step in exploring the effects of contextual information in explanations in RSs. 

We now present perspectives for future work: (1) Extend our work to other application domains, such as restaurant recommendations, tour recommendations, etc.; (2) We plan to extend our experiment to a wider range of population in order get more reliable conclusions; (3) Combine contextual information and other information such as content and demographic information to serve as explanations and explore how the they can influence the effects of explanations. 
\bibliographystyle{splncs04}
\bibliography{references} 
\end{document}